\documentclass[conference]{IEEEtran}

\usepackage{amsmath,graphicx,psfrag,multirow,lscape}
\usepackage{amssymb, amsfonts, epsfig, latexsym, times}
\usepackage{graphics,cite}
\usepackage{booktabs,balance}
\usepackage{comment}

\begin{document}
\title{Finite-State Markov Modeling of Tunnel Channels in Communication-based Train Control (CBTC) Systems}

\author{
\IEEEauthorblockN{Hongwei Wang$^{\ddag}$,~F. Richard Yu$^{\dag}$, Li Zhu$^{\ddag}$, Tao Tang$^{\ddag}$, and Bin Ning$^{\ddag}$}\\
$^{\dag}$Department of Systems and Computer Engineering, Carleton University, Ottawa, ON, Canada\\
$^{\ddag}$State Key Laboratory of Rail Traffic Control and Safety, Beijing Jiaotong University, P.R. China
%\IEEEauthorblockA{Email:basil19851028@gmail.com, richard\_yu@carleton.ca, zhulibjtu@gmail.com, tbning@bjtu.edu.cn,}

}

%\author{\IEEEauthorblockN{Qingmin~Wang$^{\dag}$, F.~Richard~Yu$^{\ddag}$, and Yi~Sun$^{\dag}$}
%\IEEEauthorblockA{$^{\dag}$School of Information and Communication Engineering, Dalian University of Technology, Dalian, China\\
%$^{\ddag}$Department of Systems and Computer Engineering, Carleton University, Ottawa, ON, K1S 5B6, Canada\\
% Email: anna.wangqingmin@gmail.com; Richard\_Yu@carleton.ca; lslwf@dlut.edu.cn
%} }

\maketitle

\begin{abstract}
Communication-based train control (CBTC) is gradually adopted in urban rail transit systems, as it can significantly enhance railway network efficiency, safety and capacity. Since CBTC systems are mostly deployed in underground tunnels and trains move in high speed, building a train-ground wireless communication system for CBTC is a challenging task. Modeling the tunnel channels is very important to design and evaluate the performance of CBTC systems. Most of existing works on channel modeling do not consider the unique characteristics in CBTC systems, such as high mobility speed, deterministic moving direction, and accurate train location information. In this paper, we develop a finite state Markov channel (FSMC) model for tunnel channels in CBTC systems. The proposed FSMC model is based on real field CBTC channel measurements obtained from a business operating subway line. Unlike most existing channel models, which are not related to specific locations, the proposed FSMC channel model takes train locations into account to have a more accurate channel model. The distance between the transmitter and the receiver is  divided into intervals, and an FSMC model is applied in each interval. The accuracy of the proposed FSMC model is illustrated by the simulation results generated from the model and the real field measurement results.
% Due to the wide availability of commercial-off-the-shelf WLAN (Wireless Local Area Network) equipment and open IEEE $802.11$ standards, WLAN-based CBTC is gaining popularity in urban rail transit systems.
%Train ground communication subsystem is the key part of CBTC systems, which often applies Wireless Local Area Networks (WLANs) as the main method. As urban rail transit systems are mostly deployed in underground tunnels and trains move fast, the performance of wireless communication in tunnels needs to be studies due to the large scale fading and the small scale fading caused by lots of reflections, scattering and Doppler frequency. In this paper, based on the measurement results in the tunnel of Beijing Subway Changping Line, we propose a novel Finite State Morkov channel (FSMC) model for CBTC systems, considering both the large scale fading and the small scale fading. Akaike Information Criterion with a correction (AICc) method is applied to get the distribution of SNR from the experimental data and then Lloyd-Max is used to partition the amplitude of SNR into several regions. Through dividing the distance into many interval, we build the relationship between the FSMC model and the distance. And the contrasts between the FSMC model and the experimental data show that the distance interval and the number of states in the model both affect the accuracy of the model.
\end{abstract}



% \fi
\IEEEpeerreviewmaketitle

%Extensive numerical examples illustrate the effectiveness of the proposed scheme.

%I wish you the best of success.
\IEEEpeerreviewmaketitle

%\begin{keywords}
%CBTC, finite-state Markov chain, WLAN
%\end{keywords}

\section{Instruction}
Urban rail transit systems are developing rapidly around the world. Duo to the huge urban traffic pressure, improving the efficiency and capacity of urban rail transit systems is increasingly in demand. Being a key subsystem in rail transit systems, communications-based train control (CBTC) is an automated train control system using  bidirectional train-ground communications to ensure the safe operation of rail vehicles \cite{What_is_communication-based_train_control}. It can enhance the level of safety and service offered to customers and improve the utilization of railway network infrastructure. CBTC is a modern successor of a traditional railway signaling system using interlockings, track circuits, and signals \cite{CBTC_Standard}.

%As a result, communication-based train control (CBTC) systems have been gradually adopted in subway signaling systems. CBTC systems require each train to report its location to the control center via a data communication system (DCS) \cite{}. Train ground communication plays the key role in CBTC systems, as it can realize the real-time communication between trains and the control center \cite{}. Therefore, the headway between trains can be shortened and the efficiency can be dramatically improved.
%Nowadays, wireless local area networks (WLANs) operating at 2.4GHz are widely used as train ground communication technology due to available commercial-off-the-shelf equipment, open standards and interoperability. There are several WLAN-based CBTC systems deployed in the world, such as Las Vegas Monorail from Alcatel and Beijing Metro Yizhuang Line and Changping Line from Beijing Jiaotong University[ZhuLi].

Building a train-ground wireless communication system for CBTC is a challenging task. As urban rail transit systems are mostly deployed in underground tunnels, there are a large amount of reflections, scattering and barriers that  severely affect the propagation performance of wireless communications. Moreover, due to the available commercial-off-the-shelf equipments, wireless local area networks (WLANs) are often adopted as the main method of train ground communications for CBTC systems \cite{ZHULI_FSMC}. However, most of the current IEEE 802.11 WLAN standards are not originally designed for the high speed environment in tunnels \cite{ZHULI_FSMC,ZFBT12}. Furthermore, the fast movement of trains will cause frequent handoffs between WLAN access points (APs), which could affect CBTC performance severely.

Modeling the channels of urban rail transit systems in tunnels is very important to design and evaluate the performance of CBTC systems. There are some previous works  on radio wave propagation in tunnels \cite{ZHANGYP_Enhancement_of_rectangular_tunnel_waveguide_model}, \cite{ZhangYP_Novel_Model}. A path loss model of tunnels is given in \cite{ZhangYP_Novel_Model}, which describes the characteristics of the large scale fading. Authors of \cite{KEGUAN} illustrate a measurement method of $2.4GHz$ in a subway tunnel, and the research object is a distributed antenna system, which is not often applied in CBTC systems. Due to the good balance between accuracy and complexity, finite state Markov channel (FSMC) model has been successfully used in different channels, including Rayleigh fading \cite{Finite_state_Markov_channel_a_useful_model_for_radio_communication_channels}, Ricean fading  \cite{Finite-state_Markov_modeling_of_correlated_Rician-fading_channels} and Nakagami fading \cite{Fast_simulation_of_diversity_Nakagami_fading_channels_using_finite-state_Markov_models}.

% give the simulation results of the Rayleigh fading channel, the Rice fading channel and the Nakagami fading channel with the FSMC model and show the accuracy of FSMC models in modeling the small scale fading channels.

Although some excellent works have been done on modeling channels, most of them do not consider the unique characteristics in CBTC systems, such as high mobility speed, deterministic moving direction,  and accurate train location information. In this paper, we develop a finite state Markov channel model  for  tunnel channels in CBTC systems. Some distinct features of the proposed channel model are as follows.
\begin{itemize}
\item The proposed FSMC model is based on real field CBTC channel measurements obtained from business operating Beijing Subway Changping line.

\item Unlike most existing channel models, which are not related to specific locations, the proposed FSMC channel model takes train locations into account to have a more accurate channel model.

% WLANs for CBTC systems according to the measurement results in the field tests with the wireless network configuration as same as the practical subway lines, which is beneficial for the practical engineering.

\item The distance between the transmitter and the receiver is divided into intervals, and an FSMC model is applied in each interval.

%As the amplitude range of signal-to-noise-ratio (SNR) for each interval can be determined by the large scale fading, then the relationship between the FSMC model and the distance can be built.

\item Lloyd-Max technique \cite{Least_squares_quantization_in_PCM} is used to determine the SNR level boundaries in the proposed FSMC model.

\item The accuracy of the proposed FSMC model is illustrated by the simulation results generated from the model and the real field measurement results. The effects of different parameters are also discussed.

%\item
% The FSMC model is compared with the experimental data, and it shows that the length of distance interval and the number of the states can affect the accuracy of the model.
\end{itemize}

%In the paper, we will propose a novel FSMC channel model of tunnels based on measurements and show how the distance affects the accuracy of the model, which is suitable for CBTC application scenarios.
%The measurement zone is divided into many intervals and we partition the SNR into different regions with the LLOYD-MAX algorithm according to the amplitudes of SNR in each distance interval. Based on the state probabilities and the transition probabilities, the simulation samples can be generated which can be verified through the experimental results.

The rest of this paper is organized as follows. Section \ref{OverviewCBTC} describes an overview of CBTC systems. In Section \ref{Sec_Measurement}, the real field measurement configuration and scenario are described. In Section \ref{Sec_FSMC}, the FSMC model is introduced. Then, Section \ref{Sec_TestandDiscussion} presents the real field measurement results and discussions. Finally, the paper is concluded in Section \ref{Sec_CandF} with future work.

\section{Overview of Communication-Based Train Control}
\label{OverviewCBTC}
%Fig. \ref{CBTC} describes a CBTC system.
In CBTC systems, continuous bidirectional wireless communications between each mobile station (MS) on the train and the wayside access point (AP) are adopted instead of the traditional fixed-block track circuit. The railway line is usually divided into areas or regions. Each area is under the control of a zone controller (ZC) and has its own radio transmission system. Each train transmits its identity, location, direction and speed to the ZC. The radio link between each train and the ZC should be continuous so that the ZC knows the locations of all the trains in its area at all the time in order to guarantee train operation safety and efficiency.

%The ZC transmits to each train the location of the train in front of it and gives it a braking curve to enable it to stop before it reaches that train. Theoretically, as long as each train is traveling at the same speed and they all have the same braking capability, they can travel together as close as a few meters in between them. When a train moves away from the coverage of an AP and enters the coverage of another AP along the railway, the handoff procedure may result in communication interruption and long latency. In CBTC systems, it is important to maintain communication link availability

Wireless channels in CBTC systems are different from those in other wireless systems, since most CBTC systems are deployed in underground tunnels, where there are a large amount of reflections, scattering and barriers that  severely affect the propagation performance of wireless communications. In order to design and evaluate the performance of CBTC systems,  modeling of tunnel channels in CBTC systems should be carefully studied.

\section{Real Field CBTC Channel Measurements}
\label{Sec_Measurement}
The objective of the real field CBTC channel measurements  is to get the real field data of WLAN propagation in tunnels under real conditions of the subway line, which can be used to build an FSMC model.

%With this objective, the preparation of the measurements consists of two parts as follows.
%\begin{enumerate}
%\item
%We need to make sure that the configuration of the measurements is the  same as business operating subway lines, including the choice of antennas, the location of antennas, and the settings of the transmitter and the receiver.
%
%\item
%We need to develop a measurement method to map channel data, including the signal strength and SNR, to the location of the receiver, which will be used in our research  that takes train locations into account to have a more accurate channel model.
%
%\end{enumerate}

%\begin{figure}[tp]
%\begin{center}
%  \includegraphics[width=0.48\textwidth]{CBTC.eps}
%  \caption{A communication-based train control (CBTC) system.}\label{CBTC}
%\end{center}
%\end{figure}

\subsection{Measurement Equipment}
Two sets of Cisco 3200 are used, and one is set as AP while the other one is set as the mobile station (MS). Both of them are set to work at the frequency of 2.412$GHz$, which is also called channel $1$. The output power of the AP is set as 30$dBm$. The AP is located on the wall of the  tunnel,
%  and the interval of APs is different in different environmental conditions, such as straight tunnels and curving tunnel,
while the MS is located on the measurement vehicle. %In our measurement, test antennas are just the same as those applied in real subway lines in operation in order to get the accurate data and achieve the detailed performance description of CBTC train-ground communication.
The transmitting antenna is a Yagi antenna connected with the AP, which is directional and vertically polarized. The half power beam width (HPBW) is  $30 ^{\circ}$ and the gain of Yagi antenna is 13.5$dBi$. In addition, the Shark-fin antenna is applied as the receiving antenna connected with the MS, which is also directional and vertically polarized. The HPBW is  $40 ^{\circ}$ and the gain of Shark-fin antenna is 10$dBi$.The measurement configuration settings are shown in Table \ref{Mconfig}.

%The antenna patterns of both antennas are shown in Fig. \ref{antenna_pattern}.

\begin{table}[tp]
\caption{Measurement Configuration}
\label{Mconfig}
\begin{tabular}{|c|c|c|}
\hline
Frequency & \multicolumn{2}{c}{$2.412GHz$} \vline \\ \hline
Transmitting Power   &   \multicolumn{2}{c}{$30dBm$} \vline \\  \hline
\multirow{4}*{Transmitting Antenna} &Type           & Yagi Antenna \\ \cline{2-3}
                                    &Polarization Direction & Vertical \\ \cline{2-3}
                                    &Gain                  & $13.5 dBi$ \\ \cline{2-3}
                                    &HPBW                  &$30 ^{\circ}$ \\ \hline
\multirow{4}*{Receiving Antenna }  &Type           & Shark-fin Antenna \\ \cline{2-3}
                                    &Polarization Direction & Vertical \\ \cline{2-3}
                                    &Gain                  & $10 dBi$ \\ \cline{2-3}
                                    &HPBW                  &$40 ^{\circ}$ \\ \hline
\end{tabular}

\end{table}

%\begin{figure}[tp]
%\centering
%\includegraphics[width=3.2in]{antenna_pattern.jpg}
%\caption{Antenna patterns of the Yagi Antenna and Shark-fin Antenna.}
%\label{antenna_pattern}
%\end{figure}

%There is a software to record the received signal strength of WGB every 200ms.
The location of the train is obtained through a velocity sensor installed on the wheel of the measurement vehicle, which can detect the realtime velocity, and the resolution of position is millimeter per second.
%The measurement principle is shown in Fig. \ref{TP}.
When the measurement vehicle is moving, the velocity sensor gets the speed transmitted to the singlechip computer immediately through a serial port. At the same time, the MS captures the signal strength and SNR at the current position, and the signal information together with the integrated displacement data by the singlechip computer can be stored in the laptop. Therefore, the signal strength and SNR mapping with the location of receiver can be obtained, which is useful to build an FSMC model depending on the distance between the transmitter and the receiver.
%\begin{figure}[tp]
%\centering
%\includegraphics[width=2.5in]{Vsensor.jpg}
%\caption{A velocity sensor.}
%\label{Vsensor}
%\end{figure}
  %There is a singlechip to control the velocity sensor and transferred the velocity to displacement. Then the position information can be transmitted through the serial port of the singlechip controller while WLANtest is recording the signal strength.%
%Then we can get the one-to-one mapping of signal strength, SNR and the exact position, which can assist in concluding the characteristics of WLAN propagation in tunnels.

%\begin{figure}[tp]
%\centering
%\includegraphics[width=3.3in]{2TestPrinciple.eps}
%\caption{The measurement principle.}
%\label{TP}
%\end{figure}

\subsection{Measurement Scenario}
\begin{figure}[tp]
\centering
\includegraphics[width=0.45\textwidth]{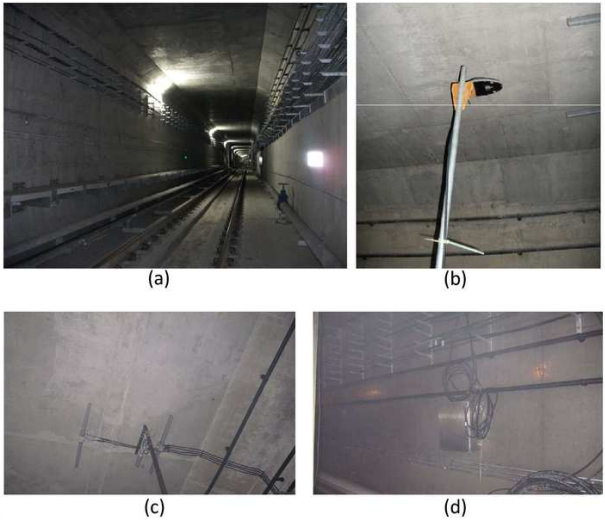}
\caption{(a) The tunnel where we performed the measurements in Beijing Changping Subway Line. (b) The Shark-fin antenna. (c) The Yagi antenna. (d) The AP set on the wall.}
\label{RT}
\end{figure}

The measurement was performed in the straight section of tunnels in Beijing Subway Changping Line, and the cross section of tunnel is rectangular. The height of the tunnel is 4.91m and the width is 4.4m. The transmitting antenna is located 0.15m below the tunnel roof, which is 4.76m. The receiving antenna is set on the top of an iron bar, which is 3.8m and also the height of the top of the train . As the threshold of the receiver is $-90dBm$, the coverage of one AP is about $0m$-$500m$, which is also the experimental zone in our measurements. The tunnel where we performed the measurement is a section of straight tunnel, and Fig. \ref{RT} shows the cross section of tunnel in Changping Subway Line, the Shark-fin antenna, the Yagi antennas and the AP set on the wall.

%\begin{figure}[tp]
%\centering
%\includegraphics[width=2.7in]{TS.jpg}
%\caption{The tunnel section and antennas deployment.}
%\label{TS}
%\end{figure}

%\begin{figure}[tp]
%\centering
%\includegraphics[width=0.48\textwidth]{RouteMapTunnel.jpg}

%\caption{The route map of the tunnel where we performed the measurements.}
%\label{RouteMap}
%\end{figure}

%For convenience, we take the center of the tunnel as the original point, then the locations of the transmitting antenna and the receiving antenna could be determined, $(0,2.3050)$ and $(0,1.8450)$, respectively.
%We pushed the measurement vehicle slowly at the speed of about 0.35m/s while doing experiments, and then much enough data can be captured.
%The purpose of our test is to get the signal strength received by WGB via the distance between the receiving antenna and the transmitting antenna, whose

\section{The Finite-State Markov Chain Channel Model}
\label{Sec_FSMC}
To capture the characteristics of tunnel channels in CBTC systems, we define channel states according to the different received SNR levels, and use an FSMC to track the state variation. In this section, we first describe the FSMC model, followed by the determination of key model parameters, including  SNR levels and SNR distribution.

\subsection{The Finite State Markov Channel Model}
Let $\Gamma$ denote the SNR of the received signal, whose range can be obtained from the experimental data. The range of SNR is partitioned into $N$ non-overlapping levels with thresholds $\{\Gamma_{n}, n = 1, 2, 3, ..., N+1\}$. Let $\textbf{S}=\{s_{1}, s_{2}, ..., s_{n}\}$ denote the finite channel states, and the channel state is $s_{n}$ when the SNR of the received signal belongs to the range $(\Gamma_{n}, \Gamma_{n+1})$. Then $\{\textbf{S}_{n}\}$ is a  Markov process and the transition probability $p_{n,j}$ can be shown as follows, which is independent of the index $n$.
\begin{equation}
p_{n,j}=P_{r}\{\textbf{S}_{k+1}=s_{n} \mid \textbf{S}_{k}=s_{j}\},
\label{STP}
\end{equation}
where $k = 1, 2, 3, ..., $ and $n, j \in \{1 ,2, ..., N+1\}$.

According to the property of first-order Markov chain, we assume that each state can only transit to the adjacent states, which means $p_{n,j}=0$, if $\mid{n-j}\mid>1$. With the definition, we can define a $K\times K$ state transition probability matrix $\textbf{P}$ with elements $p_{n,j}$.

Due to the effect of large scale fading, the amplitude of SNR depends on the distance between the transmitter and the receiver. It is obvious that the SNR is usually higher when the receiver is close to the transmitter; while it is lower when the receiver is far away from the transmitter. As a result, the transition probability from the high received SNR state to the low received SNR state is different when the receiver is near or far away from the transmitter, which means that the Markov state transition probability is related to the location of  the receiver.
%However, from the analysis of the experimental results, due to the effect of large scale fading and small scale fading, the characteristics of channel fading are not constant and varying based on position. It is obvious that the transition probability at the center of the coverage of AP is not the same as that at the edge of the coverage.
Therefore, only one state transition probability matrix, which is independent of the location of the receiver, may not generate accurate enough models to describe the tunnel channels. Thus, we divide the tunnel into $L$ intervals and one state transition probability matrix is generated for each interval. Specifically, $\textbf{P}^{l}, l \in \{1, 2, ..., L\}$, is the state transition probability matrix corresponding to the $l$th interval, and the relationship between the transition probability and the location of the receiver can be built. Then, $p_{n, j}^{l}$ is the state transition probability from state $s_{n}$ to state $s_{j}$ in the $l$th interval.

%Consequently, the state probabilities and the state transition probabilities can be defined as follows:
%\begin{equation}
%\begin{aligned}
%&p_{n}^{l}=P_{r}^{l}\{\textbf{S}_{k}=s_{n}\},\\
%&p_{n,j}^{l}=P_{r}^{l}\{\textbf{S}_{k+1}^{l}=s_{n}\mid\textbf{S}_{k}^{l}=s_{j}\},\\
%&p_{n,j}^{l}=0, \mbox{if} \mid{n-j}\mid>1,\\
%&\Sigma_{j=1}^{N+1}p_{n,j}^{l}=1, \forall k \in {1, 2, 3, ..., N+1},
%\end{aligned}
%\label{STP}
%\end{equation}
%where $p_{n}^{l}$ is the probability of state $n$.

Based on the measurement results, we need to determine the value of the state probability $p_{n}^{l}$ and the state transition probability $p_{n,j}^{l}$.

%\begin{equation}
%p_{n}^{l}=\frac{a^{l}_{n}\{\textbf{S}_{k}^{l}=s_{n}\}}{\Sigma_{n=1}^{N}a^{l}_{n}\{\textbf{S}_{k}^{l}=s_{n}\}},
%\label{SP}
%\end{equation}
%where $a^{l}_{n}\{\textbf{S}_{k}=s_{n}\}$ is the number of times state $s_{n}$ appears in the $l$th interval.
%% $A^{l}$ is the total number of samples in $l$th interval.
%
%\begin{equation}
%p_{n,j}^{l}=\frac{a^{l}_{n,j}\{\textbf{S}_{k+1}^{l}=s_{n}\mid \textbf{S}_{k}^{l}=s_{j}\}}{\Sigma_{n=1}^{N}\Sigma_{j=1}^{N}a^{l}_{n,j}\{\textbf{S}_{k+1}^{l}=s_{n}\mid \textbf{S}_{k}^{l}=s_{j}\}},
%\label{STP}
%\end{equation}
%where $a^{l}_{n,j}\{\textbf{S}_{k+1}^{l}=s_{n}\mid \textbf{S}_{k}^{l}=s_{j}\}$ is the number of times that state $s_{j}$ transits to state $s_{n}$ in the $l$th interval.

%%%%%%%%%%%%%%%%%%%%%%%%%%%%%%%%%%%%%%%%%%%%%%%%%%%%%% I should explain the transition probability and the distance.

\subsection{Determine the SNR Level Thresholds of the FSMC Model}
%As each $\textbf{P}^{l}$ describes the Markovian characteristics of the $l$th interval, to get the most appropriate $L$ is very important. In fact, it means to determine the length of each interval. In fact, there are many methods to partition the magnitude of SNR,
As mentioned above, getting the thresholds of SNR levels is the key factor that affects the accuracy  of the FSMC model. There are many methods to select the SNR level boundaries, and the equiprobable partition method is frequently used in previous works \cite{Finite_state_Markov_channel_a_useful_model_for_radio_communication_channels,Finite-state_Markov_modeling_of_correlated_Rician-fading_channels,Fast_simulation_of_diversity_Nakagami_fading_channels_using_finite-state_Markov_models}. As nonuniform amplitude partitioning may be useful to obtain more accurate estimates of system performance measures \cite{Finite-state_Markov_modeling_of_fading_channels_a_survey_of_principles_and_applications}, we choose the Lloyd-Max technique \cite{Least_squares_quantization_in_PCM} instead of the equiprobable method to partition the amplitude of SNR in this paper. Lloyd-Max is an optimized quantizer, which can decrease the distortion of scalar quantization. Lloyd-Max can realize uniform scalar quantization and non-uniform scalar quantization, and the latter one is used in this paper to divide the amplitude range of SNR.

Firstly, a distortion function $D$ is defined as follows.
\begin{equation}
D=\sum^{N+1}_{k=2}\int ^{x_{k}}_{x_{k-1}}f(\tilde{x}_{k}-x)p(x)dx,
\label{LLM1}
\end{equation}
where $x_{k}$ is the threshold of the $k$th SNR level, $f(x)$ is the error criterion function, and $p(x)$ is the probability distribution function of SNR.

%Then, the necessary conditions for minimum distortion are obtained by differentiating $D$ with respect to ${x_{k}}$ and ${\tilde{x_{k}}}$.

%The result of this minimization is a pair of equations \cite{Digital_communications}.
%\begin{eqnarray}
%%\begin{aligned}
%&f(\tilde{x}_{k}-x_{k})=f(\tilde{x}_{k+1}-x_{k}),\\
%\label{LLM2}
%&\int ^{x_{k}}_{x_{k-1}}f'(\tilde{x}_{k}-x)p(x)dx=0.
%%\end{aligned}
%\label{LLM3}
%\end{eqnarray}
%
%In the paper, we partition the amplitude of SNR into $N$ levels and there are $N+1$ corresponding thresholds $\{\Gamma_{n}, n= 1, 2, 3, ..., N+1\}$. Generally, the first and last thresholds are known, which are denoted by the minimum and maximum measurement values of SNR. Furthermore, the Lloyd-Max algorithm is used to divide $2^{n}$ levels, which means $N=2^{k}, k = 1, 2, 3, ....$, and $N$ is an even number. As a result, since $\Gamma_{1}$ and $\Gamma_{N+1}$ are known, $\Gamma_{\frac{N+2}{2}}$ can be obtained based on formula (\ref{LLM1})-(\ref{LLM3}). And then, $\Gamma_{\frac{N+4}{4}}$ and $\Gamma_{\frac{3N+4}{4}}$ can also be calculated according to the new variable $\Gamma_{\frac{N+2}{2}}$ when $k$ is larger than $2$. With the process being repeated, all elements of $\{\Gamma_{n}\}$ could be obtained.

 The error criterion function $f(x)$ is often taken as $x^{2}$ \cite{Digital_communications}. As a result, Then, the necessary conditions for minimum distortion are obtained by differentiating $D$ with respect to ${x_{k}}$ and ${\tilde{x_{k}}}$ as follows.
\begin{eqnarray}
\label{LLM4}
&x_{k}=\frac{\tilde{x}_{k}+\tilde{x}_{k+1}}{2}.\\
\label{LLM5}
&\int^{x_{k}}_{x_{k-1}}(\tilde{x}_{k}-x)p(x)dx=0.
\end{eqnarray}

Therefore, all elements of ${\Gamma_n}$ can be obtained according to \eqref{LLM5}. Combined with \eqref{LLM4}, the value of ${\Gamma_n}$ can be updated until the value of $D$ is the minimum, and the optimal thresholds of the SNR levels can be got.  As $p(x)$ is still not determined,  we should discuss the distribution of SNR according to the experimental sampling data, which is the last step to obtain the thresholds of SNR regions.

%Hence, substituting ${\Gamma_{k}}$ into (\ref{LLM4}), we can get
%\begin{equation}
%\int^{\Gamma_{b}}_{\Gamma_{a}}(\tilde{\Gamma}_{\frac{b+a}{2}}-x)p(x)dx=0,~~b>a~\mbox{and}~ b, a \in {1, 2, 3, ..., N+1},
%\label{LLM6}
%\end{equation}
%where $p(x)$ is the probability distribution function of SNR.
%
%Now, we should discuss the distribution of SNR according to the real field sampling data, which is the last step to obtain the thresholds of SNR levels.

\subsection{Determine the Distribution of SNR}

Deriving the distribution of SNR is the crucial step of partitioning the levels of SNR. In fact, there are some classic models to describe the distribution of signal strength, such as Rice, Rayleigh and Nakagami, and then the corresponding models of SNR can also be obtained \cite{Digital_Communication_over_Fading_Channels}. We firstly obtain the distribution of the signal strength in order to determine the model of SNR.

The Akaike information criterion (AIC) is adopted in this paper to get the approximate distribution model of the signal strength. The AIC is a measure of the relative goodness of fit of a statistical model. The general case of AIC is \cite{Model_selection_and_multimodel_inference}
\begin{equation}
AIC=-2\ln{L}+2\eta,
\end{equation}
where $\eta$ is the number of parameters in the statistical model, and $L$ is the maximized value of the likelihood function for the estimated model.
In fact, according to the relationship of $\eta$ and the number of samples $n$, AIC needs to be changed to Akaike information criterion with a correction (AICc) when ${n}/\eta<40$ \cite{Model_selection_and_multimodel_inference}.
\begin{equation}
AICc=AIC+\frac{2\eta(\eta+1)}{n-\eta-1}.
\label{F_AICc}
\end{equation}
AICc is adopted to estimate the model of the signal strength distribution instead of the classic AIC in the paper. In practice, one can compute AICc for each of the candidate models and select the model with the smallest value of AICc. The candidate models include Rice, Rayleigh, and Nakagami in the paper.

\begin{figure}[tp]
\centering
\includegraphics[width=0.45\textwidth]{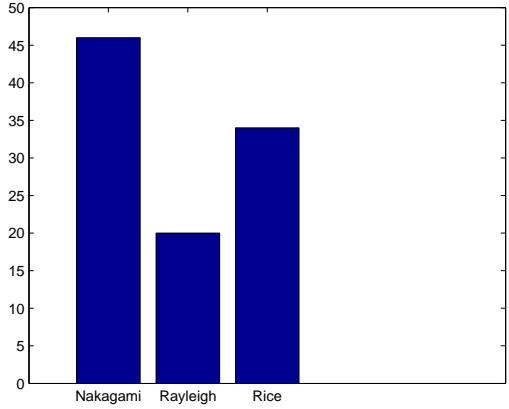}
\caption{Frequencies of AICc selecting a candidate distribution.}
\label{AICc}
\end{figure}

Our model is related to the distance between the transmitter and receiver, and the tunnel should be divided into intervals. Thus, as mentioned above, we should divide the amplitude of SNR into several levels and firstly calculate the value of AICc for each model to determine the distribution function for each interval. Now we assume there are $L$ intervals, and then we select the most appropriate model based on the frequency of the minimum AICc value of different candidate models. In order to obtain enough data for each interval, we set the length of each interval as $40$ wavelengths of WLANs \cite{A_statistical_model_for_indoor_office_wireless_sensor_channels}, and then there are $100$ intervals. Fig. \ref{AICc} shows the frequencies of AICc of different distributions. From Fig. \ref{AICc}, we can observe that the Nakagami distribution provides the best fit in a majority of the cases. As a result, we can define $p(x)$ as the Nakagami distribution.

According to \cite{Digital_Communication_over_Fading_Channels}, we can obtain the distribution of SNR,  after the distribution of the signal strength is obtained.

\begin{equation}
p(x)=\frac{m^{m}x^{m-1}}{\bar{x}^{m}\Gamma{m}}exp(-\frac{mx}{\bar{x}}),
\label{SNRfunc}
\end{equation}
where $x$ is the SNR data, $\bar{x}$ is the mean of SNR, $m$ is the fading factor of Nakagami distribution, and $\Gamma(.)$ is the gamma function. In fact, $m$ can be calculated when applying AICc through the maximum likelihood estimator for each interval.

Based on \eqref{LLM4}, \eqref{LLM5} and \eqref{SNRfunc}, the thresholds $\{\Gamma_{n},n=1,2,...,N+1\}$ of SNR in each distance interval can be derived. Table \ref{Thresholds_4} and Table \ref{Thresholds_8} demonstrate the thresholds of the SNR levels at the location of $100m$ for different intervals, where we divide SNR into four and eight levels. As the distance intervals are different, the range of SNR is different and it brings different thresholds, which can provide one more accurate model.

% After that, combined with (\ref{SP}) and (\ref{STP}), we can get the state probabilities and the state transition probabilities from the real field data.

\begin{table}[tp]
\caption{Thresholds of SNR levels ($4$ levels) at the location of $100m$ for different intervals.}
\label{Thresholds_4}
\begin{tabular}{|c|c|c|c|c|c|}
\hline
5m&10m&20m&50m&100m\\ \hline
$[95~100]$&$[90~100]$&$[80~100]$&$[50~100]$&$[0~100]$\\ \hline
24	    &22	    &22     &22     &22\\ \hline
27.98	&26.90	&27.73  &29.53	&33.22\\ \hline
32.03	&31.44	&32.89	&35.39	&44.01\\ \hline
36.31	&36.02  &38.14	&41.22	&57.00\\ \hline
41	&41     &44	    &48     &78\\ \hline
\end{tabular}
\caption{Thresholds of SNR levels ($8$ levels) at the location of $100m$ for different intervals.}
\label{Thresholds_8}
\begin{tabular}{|c|c|c|c|c|}
\hline
5m          &10m        &20m          &50m         &100m     \\ \hline
$[95~100]$&$[90~100]$&$[80~10m]$&$[50~100]$&$[0~100]$\\ \hline
24	    &22	    &22     &22     &22\\ \hline
25.99	&24.53	&25.00  &26.16	&27.87\\ \hline
27.98	&26.90	&27.73	&29.50	&33.22\\ \hline
29.99	&29.18  &30.33	&32.50	&38.50\\ \hline
32.03	&31.43  &32.89  &35.38  &44.01\\ \hline
34.13	&33.69	&35.47  &38.25	&50.04\\ \hline
36.31	&36.01	&38.14	&41.22	&57.00\\ \hline
38.59	&38.43  &40.95	&44.41	&65.68\\ \hline
41	    &41     &44	    &48     &78   \\ \hline
\end{tabular}
\end{table}

\section{Real Field Measurement Results and Discussions}
\label{Sec_TestandDiscussion}
In this section, we compare our FSMC model with real field test results to illustrate the accuracy  of the model. The effects of different parameters in the proposed model are discussed. The number of states in our model is first set as $4$. We also use $8$ states to study the effects of the number of states on the  accuracy of the proposed model. In order to obtain the effects of distance intervals on the model, we choose the intervals as $5m$, $10m$, $20m$, $50m$, $100m$.

\begin{figure}[tp]
\centering
\includegraphics[width=0.45\textwidth]{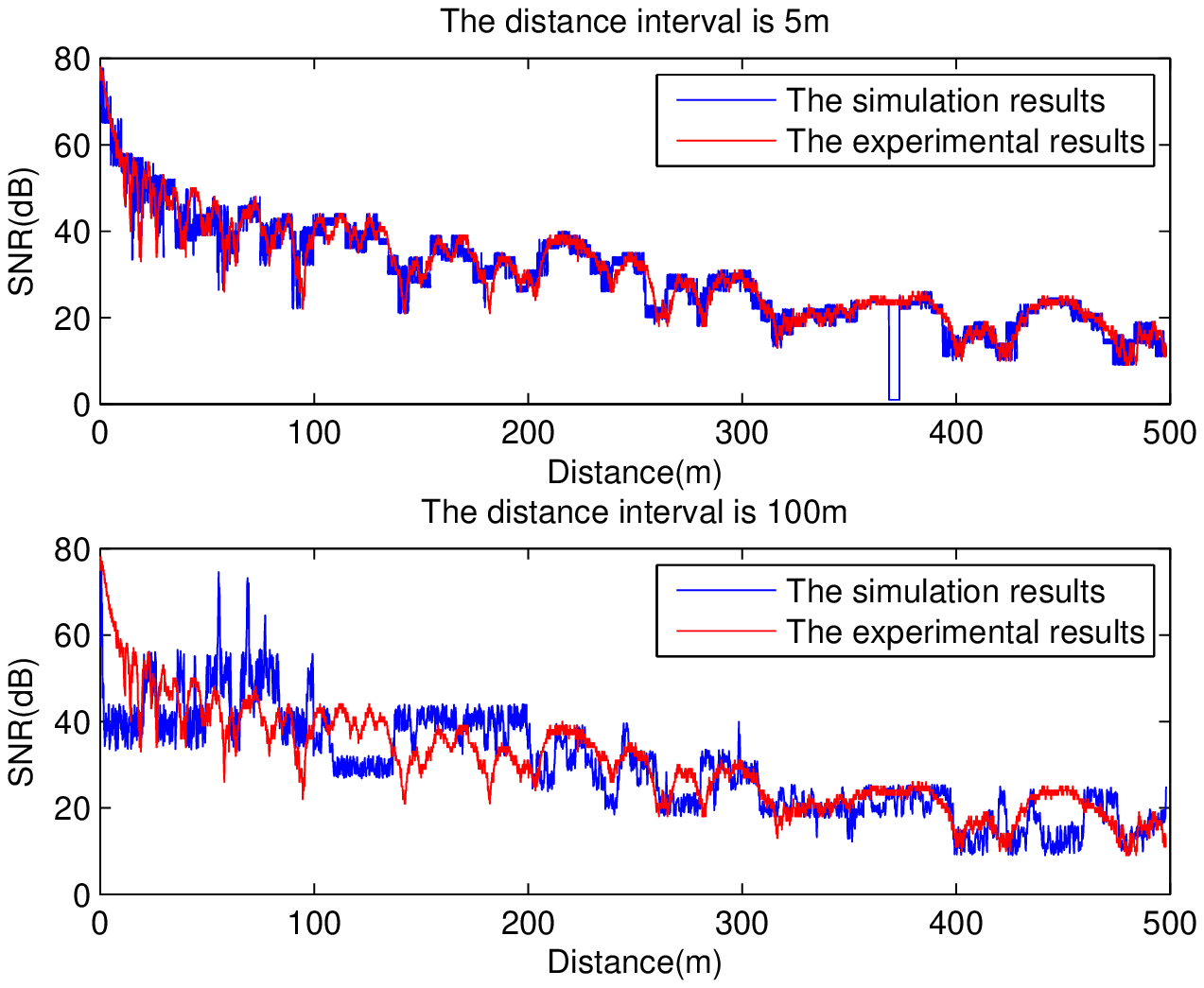}
\caption{Simulation results from the FSMC model with $5m$ and $100m$ distance interval versus experimental results from real field measurements.}
\label{5m}

\centering
\includegraphics[width=0.45\textwidth]{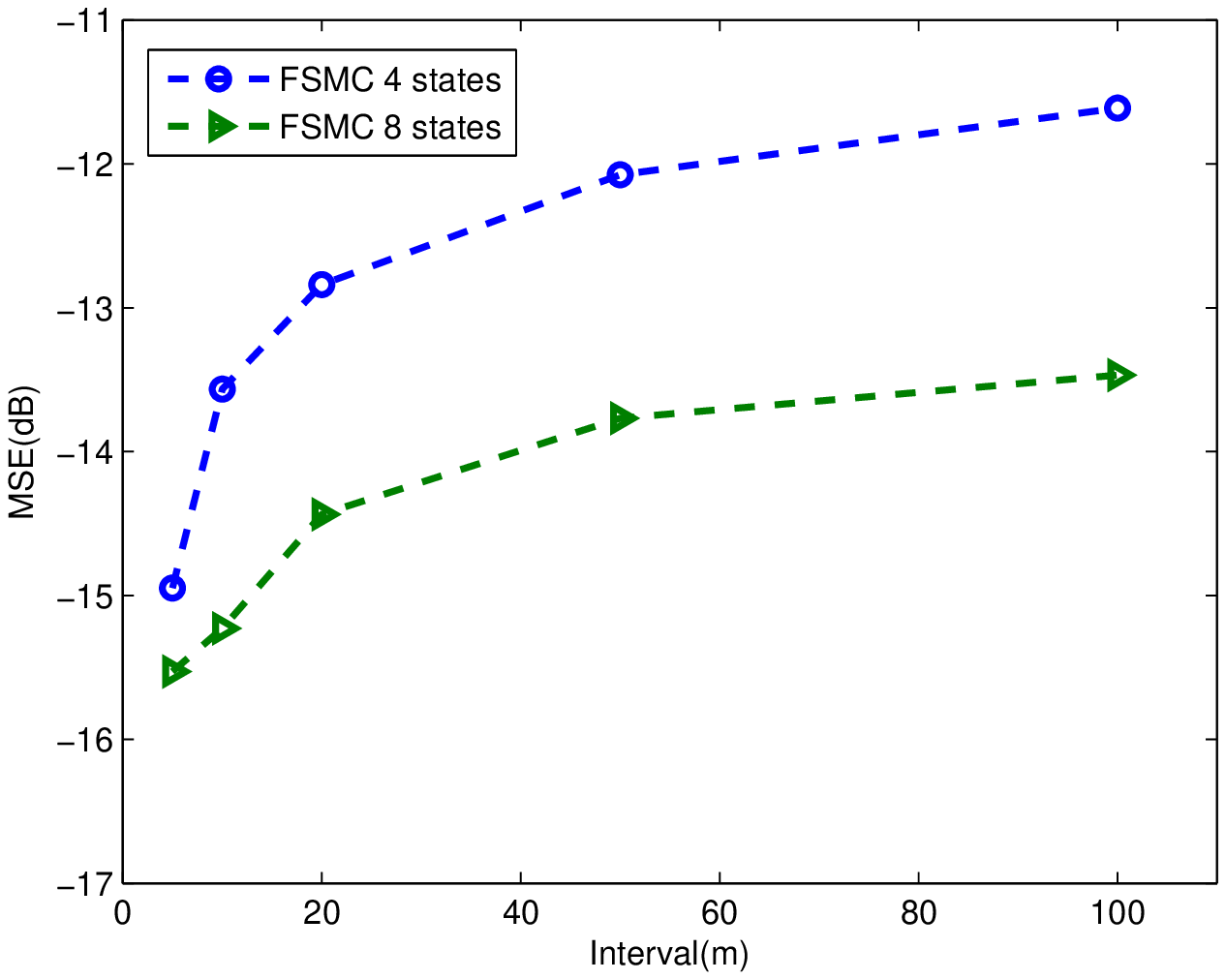}
\caption{The MSE between the FSMC model and the experimental data with $4$ states and $8$ states.}
\label{MSE}
\end{figure}

We perform the measurements in the  tunnels of Beijing Subway Changping Line many times so that enough data can be captured. We verify the accuracy of the FSMC model through another set of measurement data. First of all, we get the statistical state transition probabilities. Table \ref{STPM_8} illustrates the state transition probabilities of the FSMC model and the measurement data at the same location $(35m-40m)$ when there are eight states, and the distance interval is $5$m. Fig. \ref{5m} shows the simulation results generated from our FSMC model and the experimental results from real field measurements. We can observe that there is greater agreement between them when the distance is $5m$ than that with the $100m$ distance interval. Next, we derive the Mean Square Error (MSE) to measure the degrees of approximation, shown in Fig. \ref{MSE}. With the distance interval increasing, the MSE does also increase, which means the accuracy decreases. Moreover, it is obvious that the MSE of the FSMC model with $4$ states is larger than that with $8$ states. The number of states in the FSMC model plays a key role in the accuracy. Nevertheless, when the distance interval is $5m$, the difference of MSE is small for $4$ states FSMC model and $8$ states FSMC model. From this figure, we can see that the FSMC model with 4 states and $5m$ distance interval can provide an accurate enough channel model for tunnel channels in CBTC systems.

\begin{table}[tp]
%\caption{The values of transition probabilities for the FSMC model with $4$ states and $5m$ interval}
%\label{STPM_4}
%\begin{tabular}{|c|c|c|c|c|c|c|}
%\hline
%   &\multicolumn{3}{c}{The FSMC Model}\vline&\multicolumn{3}{c}{The Measurement Data}\vline\\ \cline{2-7}
%   &$p_{k,k-1}$&$p_{k,k}$&$p_{k,k+1}$&$p_{k,k-1}$&$p_{k,k}$&$p_{k,k+1}$\\ \hline
%%k=1&-&0.916667&0.0833333&-&0.917363&0.082637\\\hline
%%k=2&0.043478&0.869565&0.086957&0.040976&0.867864&0.09166\\\hline
%%k=3&0.024096&0.855422&0.120482&0.02488&0.85855&0.11657\\\hline
%%k=4&0.0234747&0.960094&-&0.023743&0.0976232&-\\\hline
%k=1&-&0.91&0.08&-&0.91&0.08\\\hline
%k=2&0.043&0.86&0.086&0.041&0.86&0.09\\\hline
%k=3&0.024&0.85&0.12&0.024&0.85&0.11\\\hline
%k=4&0.023&0.96&-&0.023&0.097&-\\\hline
%\end{tabular}

\caption{The values of transition probabilities for the FSMC model with $8$ states and $5m$ interval}
\label{STPM_8}
\begin{tabular}{|c|c|c|c|c|c|c|}
\hline
   &\multicolumn{3}{c}{The FSMC Model}\vline&\multicolumn{3}{c}{The Measurement Data}\vline\\ \cline{2-7}
   &$p_{k,k-1}$&$p_{k,k}$&$p_{k,k+1}$&$p_{k,k-1}$&$p_{k,k}$&$p_{k,k+1}$\\ \hline
k=1&-&0.75&0.25&-&0.78&0.22	\\\hline
k=2&0.25&0.5&0.25&0.269&0.47&0.26\\ \hline
k=3&0.25&0.5&0.25&0.23&0.5&0.26\\\hline
k=4&0.22&0.66&0.11&0.22&0.65&0.12\\\hline
k=5&0.125&0.5&0.25&0.126&0.63&0.24\\\hline
k=6&0.095&0.81&0.048&0.089&0.86&0.049\\\hline
k=7&0.13&0.6&0.27&0.12&0.61&0.26\\\hline
k=8&0.013&0.98&-&0.013&0.98&-\\\hline
\end{tabular}
\end{table}

\section{Conclusions and Future Work}
\label{Sec_CandF}
Modeling the tunnel wireless channels of urban rail transit systems is important in designing and evaluating the performance of CBTC systems. In this paper, we have proposed an FSMC model for tunnel channels in CBTC systems. Since the train location is known in CBTC systems, the proposed FSMC channel model takes train locations into account to have a more accurate channel model. The distance between the transmitter and the receiver is divided into intervals, and an FSMC model is designed in each interval. The accuracy of the proposed model has been illustrated by the simulation results generated from the proposed model and the real field measurement. In addition,  we have shown that the  number of states and the distance interval have impacts on the accuracy of the proposed FSMC model. Future work is in progress to study the effects of  wireless channels on the control performance of CBTC systems based on the proposed channel model.

\section*{Acknowledgement}This paper was supported by grants from the National Natural Science Foundation of China (No.61132003), the National High Technology Research and Development Program of China (863 Program) (2011AA110502), and projects (No. RCS2011ZZ007, RCS2012ZQ002, 2013JBM124, 2011JBZ014, RCS2010ZZ003, RCS2012K010).

\bibliographystyle{IEEEtran}
\bibliography{ref}

\begin{thebibliography}{10}
\providecommand{\url}[1]{#1}
\csname url@samestyle\endcsname
\providecommand{\newblock}{\relax}
\providecommand{\bibinfo}[2]{#2}
\providecommand{\BIBentrySTDinterwordspacing}{\spaceskip=0pt\relax}
\providecommand{\BIBentryALTinterwordstretchfactor}{4}
\providecommand{\BIBentryALTinterwordspacing}{\spaceskip=\fontdimen2\font plus
\BIBentryALTinterwordstretchfactor\fontdimen3\font minus
  \fontdimen4\font\relax}
\providecommand{\BIBforeignlanguage}[2]{{%
\expandafter\ifx\csname l@#1\endcsname\relax
\typeout{** WARNING: IEEEtran.bst: No hyphenation pattern has been}%
\typeout{** loaded for the language `#1'. Using the pattern for}%
\typeout{** the default language instead.}%
\else
\language=\csname l@#1\endcsname
\fi
#2}}
\providecommand{\BIBdecl}{\relax}
\BIBdecl

\bibitem{What_is_communication-based_train_control}
R.~Pascoe and T.~Eichorn, ``What is communication-based train control?''
  \emph{IEEE Veh. Tech. Mag.}, vol.~4, no.~4, pp. 16 --21, Dec. 2009.

\bibitem{CBTC_Standard}
IEEE, ``Standard for communications-based train control ({CBTC}) performance
  and functional requirements,'' \emph{IEEE Std 1474.1-2004 (Revision of IEEE
  Std 1474.1-1999)}, pp. 0\_1 --45, 2004.

\bibitem{ZHULI_FSMC}
L.~Zhu, F.~Yu, B.~Ning, and T.~Tang, ``Cross-layer handoff design in
  {MIMO}-enabled {WLANs} for communication-based train control ({CBTC})
  systems,'' \emph{IEEE J.\ Sel.\ Areas\ Commun.}, vol.~30, no.~4, pp. 719
  --728, May 2012.

\bibitem{ZFBT12}
L.~Zhu, F.~R. Yu, B.~Ning, and T.~Tang, ``{Handoff Performance Improvements in
  MIMO-Enabled Communication-Based Train Control Systems},'' \emph{{IEEE
  TRANSACTIONS ON Intelligent Transportation Systems}}, vol.~13, no.~2, pp.
  582--593, 2012.

\bibitem{ZHANGYP_Enhancement_of_rectangular_tunnel_waveguide_model}
Y.~Zhang and Y.~Hwang, ``Enhancement of rectangular tunnel waveguide model,''
  in \emph{Proc. APMC '97}, vol.~1, Dec. 1997, pp. 197 --200.

\bibitem{ZhangYP_Novel_Model}
Y.~Zhang, ``Novel model for propagation loss prediction in tunnels,''
  \emph{IEEE Trans.\ Veh.\ Tech.}, vol.~52, no.~5, pp. 1308 -- 1314, Sep. 2003.

\bibitem{KEGUAN}
K.~Guan, Z.~Zhong, J.~Alonso, and C.~Briso-Rodriguez, ``Measurement of
  distributed antenna systems at 2.4 ghz in a realistic subway tunnel
  environment,'' \emph{IEEE Trans.\ Veh.\ Tech.}, vol.~61, no.~2, pp. 834
  --837, Feb. 2012.

\bibitem{Finite_state_Markov_channel_a_useful_model_for_radio_communication_channels}
H.~S. Wang and N.~Moayeri, ``Finite-state {Markov} channel-a useful model for
  radio communication channels,'' \emph{IEEE Trans.\ Veh.\ Tech.}, vol.~44,
  no.~1, pp. 163 --171, Feb. 1995.

\bibitem{Finite-state_Markov_modeling_of_correlated_Rician-fading_channels}
C.~Pimentel, T.~Falk, and L.~Lisboa, ``Finite-state {Markov} modeling of
  correlated {Rician}-fading channels,'' \emph{IEEE Trans.\ Veh.\ Tech.},
  vol.~53, no.~5, pp. 1491 -- 1501, Sept. 2004.

\bibitem{Fast_simulation_of_diversity_Nakagami_fading_channels_using_finite-state_Markov_models}
C.~Iskander and P.~Mathiopoulos, ``Fast simulation of diversity {Nakagami}
  fading channels using finite-state {Markov} models,'' \emph{IEEE Trans.\
  Broadcasting.}, vol.~49, no.~3, pp. 269 -- 277, Sept. 2003.

\bibitem{Least_squares_quantization_in_PCM}
S.~Lloyd, ``Least squares quantization in {PCM},'' \emph{IEEE Trans.\ Inform.\
  Theory}, vol.~28, no.~2, pp. 129 -- 137, Mar. 1982.

\bibitem{Finite-state_Markov_modeling_of_fading_channels_a_survey_of_principles_and_applications}
P.~Sadeghi, R.~Kennedy, P.~Rapajic, and R.~Shams, ``Finite-state {Markov}
  modeling of fading channels - a survey of principles and applications,''
  \emph{IEEE Signal Proc. Mag.}, vol.~25, no.~5, pp. 57 --80, Sep. 2008.

\bibitem{Digital_communications}
J.~Proakis, \emph{Digital communications}.\hskip 1em plus 0.5em minus
  0.4em\relax McGraw-Hill, 1995.

\bibitem{Digital_Communication_over_Fading_Channels}
M.-S.~A. Marvin K.~Simon, \emph{Digital Communication over Fading
  Channels}.\hskip 1em plus 0.5em minus 0.4em\relax John Wiley \& Sons, 2005.

\bibitem{Model_selection_and_multimodel_inference}
A.~D. Burnham~KP, \emph{Model selection and multimodel inference: a practical
  information-theoretic approach}.\hskip 1em plus 0.5em minus 0.4em\relax
  Springer, 2002.

\bibitem{A_statistical_model_for_indoor_office_wireless_sensor_channels}
S.~Wyne, A.~Singh, F.~Tufvesson, and A.~Molisch, ``A statistical model for
  indoor office wireless sensor channels,'' \emph{IEEE Trans.\ Wireless
  Commun.}, vol.~8, no.~8, pp. 4154 --4164, Aug. 2009.

\end{thebibliography}
%


\end{document}